\documentclass[10pt,aps,prl,twocolumn,superscriptaddress]{revtex4}
\usepackage[linktocpage=true, colorlinks=true, urlcolor=blue, linkcolor=blue, citecolor=blue]{hyperref}
\usepackage{graphicx}
\usepackage{amsmath,amssymb}
\usepackage{color}
\usepackage{url}
\usepackage{hyperref}
\usepackage{cleveref}
\usepackage{xfrac}
\usepackage{upgreek}

\graphicspath{{./Images/}}

\begin{document}

\title{Choroidal vasculature imaging with laser Doppler holography}

\author{L\'eo Puyo}
\affiliation{Corresponding author: gl.puyo@gmail.com}
\affiliation{Institut Langevin. Centre National de la Recherche Scientifique (CNRS). Paris Sciences \& Lettres (PSL University). \'Ecole Sup\'erieure de Physique et de Chimie Industrielles (ESPCI Paris) - 1 rue Jussieu. 75005 Paris France}

\author{Michel Paques}
\affiliation{Centre Hospitalier National d'Ophtalmologie des Quinze-Vingts, INSERM-DHOS CIC 1423. 28 rue de Charenton, 75012 Paris France}
\affiliation{Institut de la Vision-Sorbonne Universit\'es. 17 rue Moreau, 75012 Paris France}

\author{Mathias Fink}
\affiliation{Institut Langevin. Centre National de la Recherche Scientifique (CNRS). Paris Sciences \& Lettres (PSL University). \'Ecole Sup\'erieure de Physique et de Chimie Industrielles (ESPCI Paris) - 1 rue Jussieu. 75005 Paris France}

\author{Jos\'e-Alain Sahel}
\affiliation{Centre Hospitalier National d'Ophtalmologie des Quinze-Vingts, INSERM-DHOS CIC 1423. 28 rue de Charenton, 75012 Paris France}
\affiliation{Institut de la Vision-Sorbonne Universit\'es. 17 rue Moreau, 75012 Paris France}

\author{Michael Atlan}
\affiliation{Institut Langevin. Centre National de la Recherche Scientifique (CNRS). Paris Sciences \& Lettres (PSL University). \'Ecole Sup\'erieure de Physique et de Chimie Industrielles (ESPCI Paris) - 1 rue Jussieu. 75005 Paris France}

\date{\today}

\begin{abstract}
The choroid is a highly vascularized tissues supplying the retinal pigment epithelium and photoreceptors. Its implication in retinal diseases is gaining increasing interest. However, investigating the anatomy and flow of the choroid remains challenging. Here we show that laser Doppler holography provides high contrast imaging of choroidal vessels in humans, with a spatial resolution comparable to state of the art indocyanine green angiography and optical coherence tomography. Additionally, laser Doppler holography contributes to sort out choroidal arteries and veins by using a power Doppler spectral analysis. We thus demonstrate the potential of laser Doppler holography to improve our understanding of the anatomy and flow of the choroidal vascular network.
\end{abstract}

\maketitle

\section{Introduction}
The choroid is a highly vascularized tissue supplying the retinal pigment epithelium (RPE) and photoreceptors which carries a notably higher blood flow than the retina itself~\cite{AlmBill1973}. It is gaining interest as a potential driver of a number of retinal diseases, a role which has been documented by indocyanine green angiography (ICG-A)~\cite{Bischoff1985} and optical coherence tomography (OCT)~\cite{Laviers2014, Ferrara2016}. Evidence indeed suggests that the choroid is involved in the pathogenesis of diseases such as central serous chorioretinopathy, myopic chorioretinal atrophy, and chorioretinal inflammatory diseases~\cite{Mrejen2013}. Additionally, the implication of choroidal vascular abnormalities has been suggested in age-related macular degeneration (AMD)~\cite{Mcleod2009}. Thus, a better insight into the choroid physiology in normal and pathological conditions appears to be crucial to improve our understanding of many ocular diseases.

The choroid is a very remarkable tissue, with the vascular compartment accounting for most of its volume. The normal choroidal circulation originates from the posterior ciliary arteries (PCAs), branching into short PCAs (SPCAs) that penetrate the sclera usually in the posterior pole~\cite{Hayreh2011}. The distribution of flow within the choroid itself is rather poorly known because the anatomical disposition of choroidal vessels remains difficult to document in patients. ICG-A enables imaging of choroidal vessels thanks to the penetration of infrared light through the optical barrier formed by the photoreceptor and epithelial cells. Although the clinical use of ICG-A has been recommended for a few highly selective chorioretinal disorders~\cite{Yannuzzi2011}, it is an invasive method that allows limited insights into the arteriovenous distinction in most eyes and does not allow flow analysis. Non-invasive optical methods include OCT which allows visualization of the choroid on cross-sections with a high signal to noise ratio and study of the choroidal thickness~\cite{Spaide2008}. OCT-angiography (OCT-A) is a speckle contrast imaging method based on OCT that allows for the volumetric imaging quality of OCT with a functional contrast sensitive to blood flow, and it has allowed to image the retinal vascular tissue with high spatial resolution~\cite{Spaide2018}. This led to the finding that many retinal diseases are related to an increased choroidal thickness, with vascular dilation being the primary cause of such increased thickness~\cite{Mrejen2013}. OCT based methods have been used with the purpose of imaging the choroid en-face~\cite{Choi2013, Poddar2014, Gorczynska2016, Kurokawa2017} but the technique faces issues including fringe washout due to the large choroidal blood flow and recent work suggests that a higher acquisition rate improves the quality of choroidal images~\cite{Migacz2019}. Laser speckle flowgraphy also allows en-face imaging of the choroid but details of choroidal vessels are rather limited~\cite{Sugiyama2010, Calzetti2018}. Ultimately, none of the above-mentioned techniques allows a robust arteriovenous differentiation and a quantitative flow approach. Very recently, line-scanning laser Doppler flowmetry was used successfully to reveal the choroidal vasculature~\cite{Mujat2019}, but no arteriovenous differentiation was investigated.

Laser Doppler holography (LDH) is a digital holographic method that relies on analyzing the beat frequency spectrum between a Doppler broadened beam and a monochromatic reference beam to extract a blood flow contrast~\cite{MagnainCastelBoucneau2014, Pellizzari2016, Donnarumma2016}. We have shown in humans that LDH can reveal blood flow in retinal vessels over an extended field of view with a temporal resolution down to a few milliseconds~\cite{Puyo2018}. Because digital holography allows for the reference beam to be the main source of light on the camera, ultrahigh speed measurements of the Doppler power spectral density (DPSD) can be performed over a full field. This allows to separate weakly Doppler shifted photon (from dynamic scattering and global eye movements) and strongly Doppler shifted photon (from pulsatile flow in blood vessels). Low-frequency shift contributions are discarded in a temporal Fourier transform analysis thus producing an effective selection of the photons that have been scattered by high velocity red cells. Owing to this ability to select strongly Doppler shifted light, we were able to reveal choroidal vessels where the flow is even greater than in the retina, resulting in even greater Doppler frequency shifts. In this work, we found that LDH was able to reveal choroidal vascular structures that were not observed with state of the art ICG-A and OCT instruments. Besides, we show that an additional Fourier analysis of the DPSD in choroidal vessels offers a robust segregation of vessels based on their blood flow which leads to an arteriovenous differentiation.

\section{Methods}

\begin{figure}[t!]
\centering
\includegraphics[width = 1\linewidth]{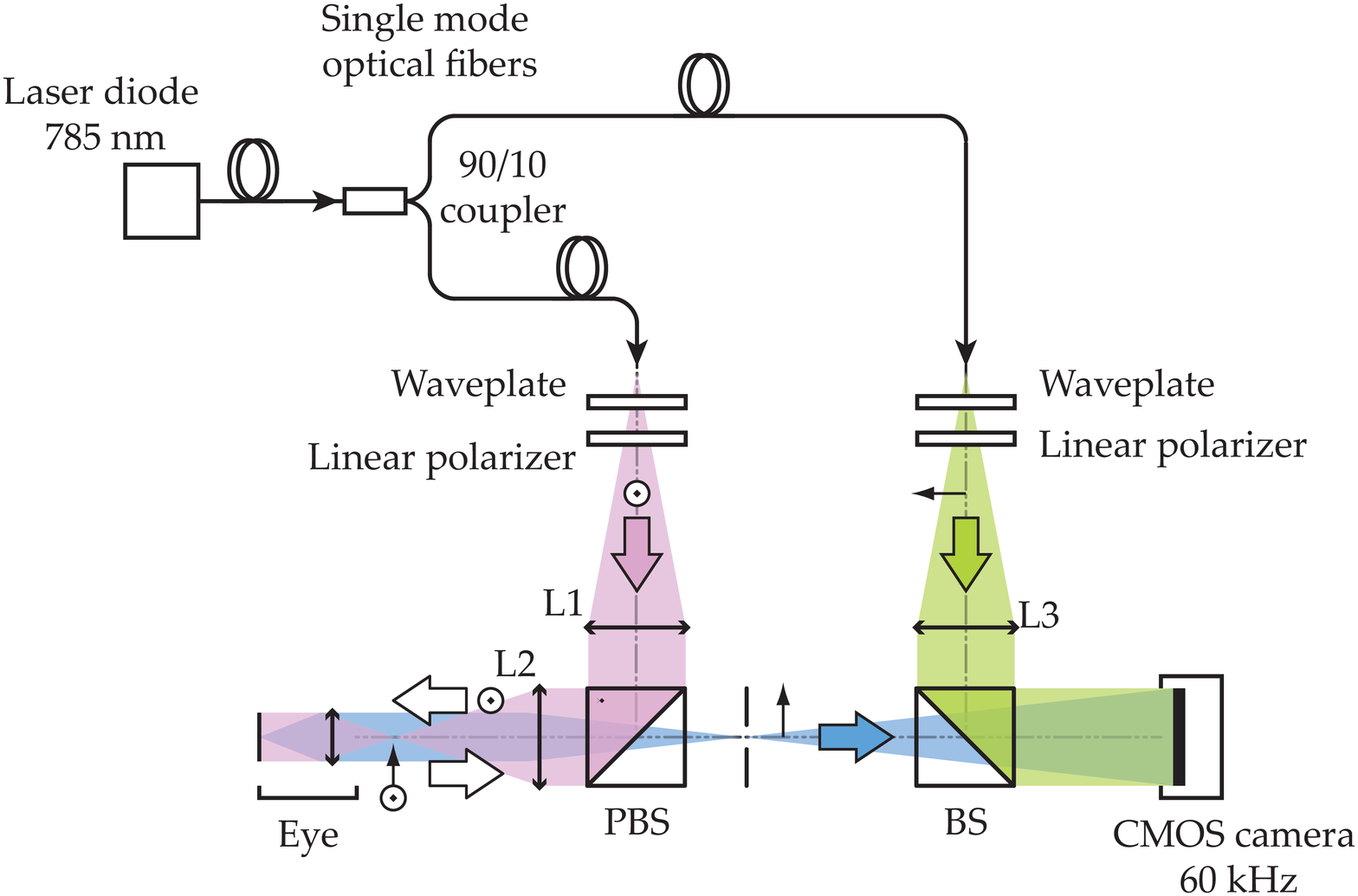}
\caption{Laser Doppler holography optical setup. L1, L2 and L3 are converging lenses. PBS: Polarizing Beam-Splitter. BS: Beam-Splitter. The light source is a single wavelength laser diode (SWL-7513-H-P, Newport). The Doppler broadened light backscattered by the retina and choroid is combined with the reference field and interferograms are recorded on the CMOS camera (Ametek - Phantom V2511) running at $60 \, \rm  kHz$.
}\label{fig_I_Setup_noaom_noslowcam}
\end{figure}

\subsection{Laser Doppler holography}
We used the laser Doppler holography setup sketched in Fig.~\ref{fig_I_Setup_noaom_noslowcam} and presented in~\cite{Puyo2018}. Briefly, it consists of a fiber Mach-Zehnder optical interferometer where the light source used for the experiments is a $45 \, \rm  mW$ and $785 \, \rm  nm$ single-mode fiber diode laser (Newport SWL-7513-H-P), spatially and temporally coherent. In contrast with our previous work, we did not use acousto-optic modulators and real-time monitoring channel which further simplifies the optical setup. The power of the laser beam incident at the cornea was $1.5 \, \rm  mW$ of constant exposure and covered the retina over approximately 4 $\times$ 4 mm$^2$. This irradiation level is compliant with the exposure levels of the international standard for ophthalmic instruments ISO 15004-2:2007. The images presented in this manuscript were obtained in the left an right eye of the same subject that were healthy except for moderate myopia. The eye used in Fig.~\ref{fig_9_HyperspectralImages} has a -3.5 diopters myopia and the eye used for all the other Figs. has -5.5 diopters myopia. Informed consent was obtained for the subject and experimental procedures adhered to the tenets of the Declaration of Helsinki.

The wave backscattered by the retina, which bears a Doppler broadened spectrum, is combined with the reference wave using a non-polarizing beamsplitter cube. The polarization of the reference wave is adjusted with a half-wave plate and a polarizer to optimize fringe contrast. The interferograms formed in the sensor plane are recorded using a CMOS camera (Ametek - Phantom V2511, quantum efficiency 40\%, 12-bit pixel depth, pixel size $28 \, \upmu \rm m$) with a $f_{S} = 60 \, \rm  kHz$ frame rate in a 512 $\times$ 512 format; the exposure time for each frame was set to $16.215 \, \upmu \rm s$. The diffracted speckle pattern is numerically propagated to the retinal plane using the angular spectrum propagation. The holographic configuration is on-axis and the reconstruction distance ($z  \approx  0.19 \, \rm  m$) is large enough so that the twin image energy is spread over the reconstructed hologram and has no appreciable effect on the resulting image.
After numerical propagation, the beat frequency spectrum of the reconstructed holograms is analyzed by short-time Fourier transform analysis which naturally removes the interferometric zero-order term. In this approach, it is considered that the cross-beating terms of the holograms carry the Doppler broadening caused by the passage of red cells and the data processing consists of measuring the local optical temporal fluctuations. Power Doppler images are calculated by integration of the DPSD over the frequency interval $[f_{\rm 1}, f_{\rm 2}]$. Power Doppler images quantitatively measure blood flow in arbitrary units that depend on both the local blood volume and blood velocity.

\subsection{Hyperspectral flow analysis} \label{subsection_MethodsHyperspectrales}

In our previous work, we analyzed the retinal blood flow changes of power Doppler images throughout cardiac cycles. In this article, we present vasculature maps obtained by averaging power Doppler images over time, and we introduce a temporal analysis method based on a Fourier segmentation of the DPSD. The chosen frequency range $[f_{\rm 1}, f_{\rm 2}]$ over which the DPSD is integrated to produce power Doppler images determines the blood vessels that are revealed according to their flow~\cite{Puyo2018}. Integrating lower Doppler frequency shifts reveal vessels with a lower blood flow while large frequency shifts reveals vessels with greater flows. When separating the low and high frequency power Doppler images and combining them into a single composite color image, we can simultaneously display vessels with a wide range of flows and qualitatively encode the flow information in the image color, as shown in subsection~\ref{subsection_HyperSpectral}.

\subsection{Stitching process} \label{subsection_HuginStitchin}
The field of view of an individual power Doppler image is about 4 $\times$ 4 mm$^2$, so in order to cover a wider field of view, we stitched together power Doppler images of multiple locations of the eye fundus. This process was used to create images displayed in subsecction~\ref{subsection_Doppler_Panorama} and \ref{subsection_HyperSpectral}. For each location, power Doppler images were averaged for a period of time approximately equal to the cardiac cycle period although for a few acquisitions, the averaging time was slightly under this duration as some frames affected by micro-saccades were discarded. The open source software Hugin was used to create the full panorama by stitching together all the individual power Doppler images. To proceed with the alignment procedure, the images were manually placed in the correct position and control points were automatically detected in manually indicated regions of interest using Hugin's CPFind native function. Hugin optimization changes the images position parameters so as to minimize the separation of the control points and improve the overall image alignment in the overlapping areas. Finally, Hugin's built-in blender compensated for the mismatches of intensity on the edges of individual images.

\section{Results}

\subsection{Choroidal vasculature around the optic nerve head} \label{subsection_ONH_Vasculature}
\begin{figure}[t!]
\centering
\includegraphics[width = 1\linewidth]{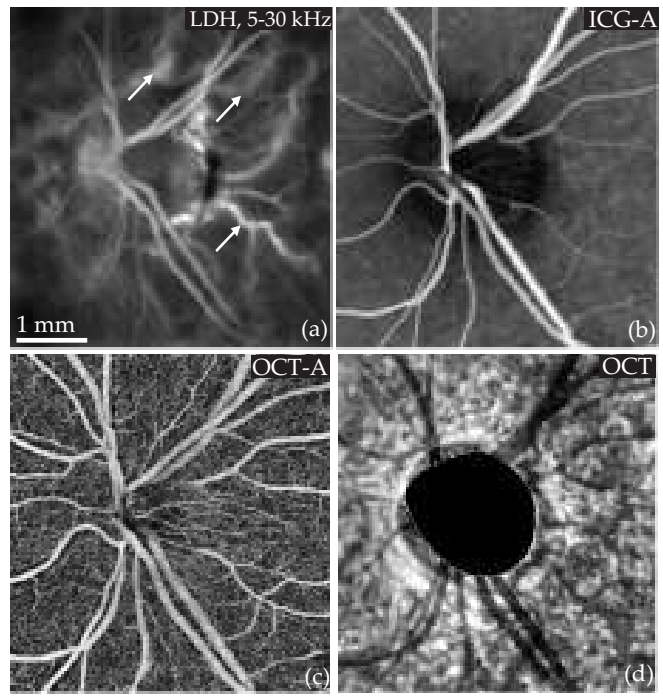}
\caption{Angiographic images of the optic nerve head (ONH) with different instruments. (a) LDH reveals the retinal vasculature and choroidal arteries (arrows) originating from paraoptic SPCAs. (b) A late ICG-A (Heidelberg - Spectralis) reveals the retinal vessels. (c) and (d): OCT-A image of the retinal layer and OCT image of the choroid without the choriocapillaris (Optovue, Avanti with AngioVue), respectively.
}\label{fig_II_Doppler_ICG_OCTA_ONH}
\end{figure}

In Fig.~\ref{fig_II_Doppler_ICG_OCTA_ONH}, we compare different angiographic instruments by imaging the optic nerve head (ONH) region in the same eye. A power Doppler image is presented in Fig.~\ref{fig_II_Doppler_ICG_OCTA_ONH}(a), and Fig.~\ref{fig_II_Doppler_ICG_OCTA_ONH}(b) shows the equivalent ICG angiogram (Heidelberg - Spectralis HRA). Figure~\ref{fig_II_Doppler_ICG_OCTA_ONH}(c) and \ref{fig_II_Doppler_ICG_OCTA_ONH}(d) show respectively an OCT-A image of the retinal layer and an OCT image of the choroid without the choriocapillaris (Optovue, RTVue XR Avanti with AngioVue). The RTVue XR Avanti that was used is a spectral-domain OCT that can acquire up to 70,000 A-Scan/s with an operating wavelength centered on $840 \, \rm  nm$.

The power Doppler image in Fig.~\ref{fig_II_Doppler_ICG_OCTA_ONH}(a) reveals the retinal vessels with a bright contrast due to LDH blood flow sensitivity. Other vascular structures around the ONH can be noticed such as choroidal arteries originating from paraoptic SPCAs in the surroundings of the optic disc (arrows). We assume the vascular structure revealed by LDH around the ONH is the peripapillary choroidal arcade, possibly made more visible by a local photoreceptor atrophy. With ICG-A, the retinal vessels and the two cilioretinal arteries are clearly revealed, but the choroidal arteries cannot be observed. The timing of the acquisition of the ICG angiogram presented in this Fig. was late enough that the dye had reached the venous retinal vasculature. However, as discussed, in the ICG-angiogram of the arterial timing in the same eye shown in subsection~\ref{subsection_ICGA_phases}, the choroidal arteries are still not revealed close to the ONH. OCT-A reveals retinal vessels with a very good sensitivity to small blood flows and allows to effectively identify to which layers belong the blood vessels. For the part of the choroid that is below the choriocapillaris, we show an OCT image instead of OCT-A as we found it revealed more efficiently the larger choroidal vessels. However even with OCT the imaging quality of these vessels is rather limited.

Hence in the images presented, LDH is able to reveal retinal vascular structures and also other deeper vascular structures that were not observed with ICG-A or with difficulty with OCT (Optovue, RTVue XR Avanti).

\subsection{Choroidal vasculature in the posterior pole} \label{subsection_Doppler_Panorama}
\begin{figure}[]
\centering
\includegraphics[width = 1\linewidth]{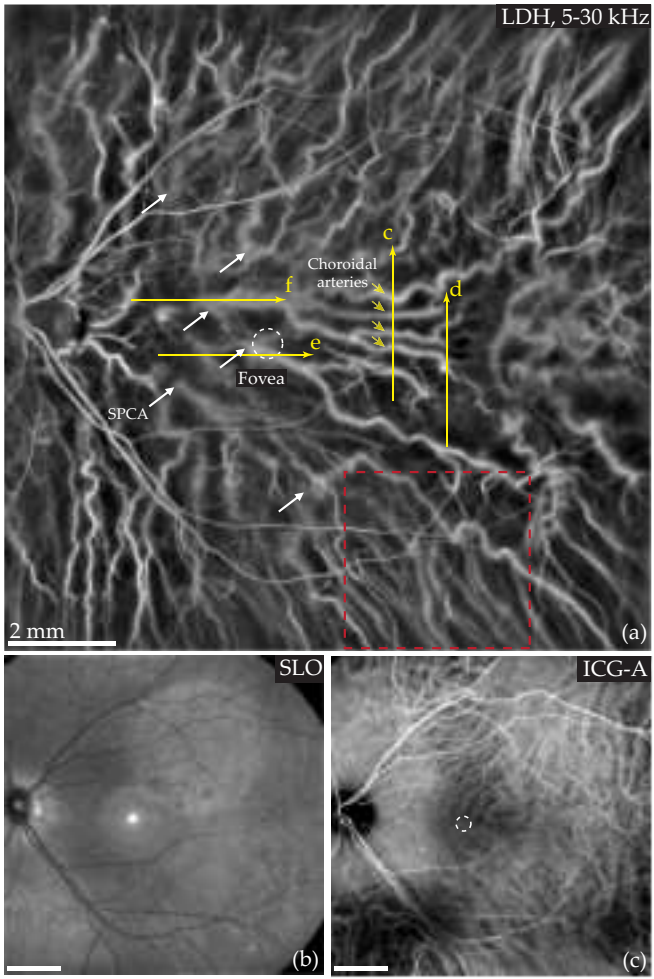}
\caption{LDH, SLO and ICG-A in the same eye. (a) 5 $\times$ 5 power Doppler images calculated over $5-30 \, \rm  kHz$ are stitched to produce a panorama on which retinal and choroidal vessels can be observed. The white circle indicates the fovea; the white arrows mark temporal distal SPCAs; the yellow lines shows the location of the B-scans presented in Fig.~\ref{fig_IV_Doppler_vs_OCTA_Macula} and the yellow arrows mark the position of the deep choroidal arteries identified in Fig.~\ref{fig_IV_Doppler_vs_OCTA_Macula}(c). Finally, the red square indicates the location of the images in Fig.\ref{fig_VI_HyperspectralImages_Spectra}. (b) SLO (Heidelberg - Spectralis) image of the retina. (c) Late ICG angiogram (Heidelberg - Spectralis) image showing retinal and choroidal vessels (mostly veins).
}
\label{fig_III_Panorama_vs_ICG}
\end{figure}
In Fig.~\ref{fig_III_Panorama_vs_ICG} a wide-field LDH panorama is compared to SLO and ICG-A. The power Doppler image shown in Fig.~\ref{fig_III_Panorama_vs_ICG}(a) was made out of 5 $\times$ 5 individual images that were stitched together following the process described in subsection~\ref{subsection_HuginStitchin}. The power Doppler images were calculated using the frequency range $[f_{\rm 1}, f_{\rm 2}] = 5-30 \, \rm  kHz$. To help with the registration all images overlapped and the total field of view is estimated at $12 \, \rm  mm$. Figure~\ref{fig_III_Panorama_vs_ICG}(b) and \ref{fig_III_Panorama_vs_ICG}(c) show the same region imaged with the SLO and ICG-A modules of a Spectralis-HRA (Heidelberg).

SLO reveals the vessels that belong to the retina, so when comparing images obtained with SLO and LDH, it is manifest that the vast majority of the vessels revealed by the latter instrument belongs to the choroid. ICG-A also reveals some choroidal vessels as well as retinal vessels, but when comparing LDH with ICG-A it is apparent that these two angiograms do not consistently show the same choroidal vessels. For example close to the macular region, ICG-A does not reveal any vessel whereas LDH shows several very large choroidal vessels. Conversely, some vessels apparent with ICG-A are not visible on this LDH panorama.

In the posterior pole, the choroidal vessels visible on power Doppler images calculated with high frequency shifts tend to be choroidal arteries while the choroidal vessels visible with this ICG angiogram are mostly veins. This conjecture is consistent with the known anatomy of the choroid, as approximately 10-20 SPCAs from the sclera penetrate the choroid around the ONH and macular region and run radially towards the equator~\cite{Hayreh2004, Anand2010}. The SPCAs then branch into choroidal arteries that supply the RPE and photoreceptors in a segmented organization~\cite{Hayreh1975}.

The type of vessels that are revealed with LDH is determined by the frequency range used to compute power Doppler images. As demonstrated in subsection~\ref{subsection_HyperSpectral}, to some extent it is possible to produce power Doppler images more similar to late phase ICG-A when using a lower frequency range. To make this panorama, the DPSD was integrated over the frequencies $5-30 \, \rm kHz$ which gives more importance to larger flows as the total power Doppler signal is naturally greater in areas with a greater flow such as arteries. However as explained in subsections~\ref{subsection_HyperSpectral} and~\ref{subsection_VortexVeins}, outside the posterior pole, choroidal arteries tend to stay visible during the late phase ICG-A while choroidal veins become visible at high Doppler frequency shifts. It can also be noticed in the bottom right part of this LDH panorama that some choroidal veins can already be observed with LDH.

To some extent there might have been variations of quality between individual acquisitions (eye position etc) which may explain the differences in contrast that can be noticed between some parts of the panorama. It is also likely that the variations in density of photoreceptors and epithelial cells has an impact on the imaging quality of the underlying vessels. In particular, in the macular region where the density of photoreceptor and epithelial cells is the highest, it seems the quality of the power Doppler images is slightly lower compared to the rest of the panorama.

\begin{figure}[t!]
\centering
\includegraphics[width = 1\linewidth]{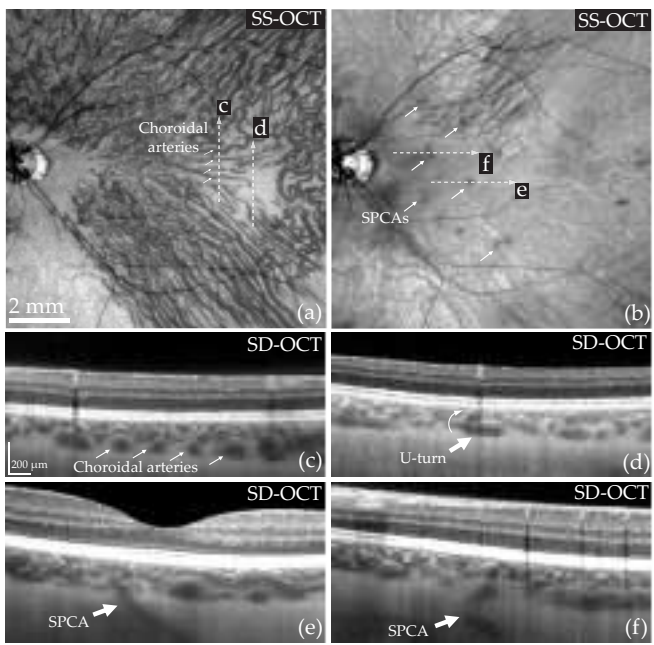}
\caption{Structural OCT images of the same eye as in Fig.~\ref{fig_III_Panorama_vs_ICG}. (a) and (b): en-face SS-OCT images (Zeiss - Plex Elite 9000) showing the deep choroid and the sclera, respectively. The arrows indicate choroidal arteries in (a) and SPCAs in (b). (c-f): SD-OCT B-scans (Heidelberg - Spectralis). (c) Cross-section showing the position of four choroidal arteries in the choroid (arrows). (d) A choroidal artery makes a U-turn to approach Bruch's membrane. (e-f) The entry points of SPCAs visible with LDH are also revealed with OCT.
}
\label{fig_IV_Doppler_vs_OCTA_Macula}
\end{figure}

We used state of the art OCT instruments in the same eye as the one used to make the images presented in Fig.~\ref{fig_III_Panorama_vs_ICG} in order to give another insight of the same choroidal structures. A swept-source OCT (Zeiss - Plex Elite 9000) was used to produce en-face images shown in Fig.~\ref{fig_IV_Doppler_vs_OCTA_Macula}(a) and \ref{fig_IV_Doppler_vs_OCTA_Macula}(b). This instrument can acquire 12 $\times$ 12 mm images with a relatively high A-scan rate (100 kHz) and wavelength of operation  ($1050 \, \rm  nm$) which allow a better penetration depth into the choroid. We also used a spectral-domain OCT (Heidelberg - Spectralis) to make cross-sectional images in areas where we identified specific choroidal structures with LDH. We present the structural rather than the OCT-A images obtained by swept-source and spectral-domain OCT (SS-OCT and SD-OCT, respectively) as we found they revealed the deeper choroidal vasculature more efficiently. Figure~\ref{fig_IV_Doppler_vs_OCTA_Macula}(a) and \ref{fig_IV_Doppler_vs_OCTA_Macula}(b) show en-face SS-OCT images of the deep choroid and sclera, and Fig.~\ref{fig_IV_Doppler_vs_OCTA_Macula}(c) to (f) are SD-OCT B-scans. The positions of all B-scans are indicated in Fig.~\ref{fig_III_Panorama_vs_ICG}, those presented in Fig.~\ref{fig_IV_Doppler_vs_OCTA_Macula}(c) and \ref{fig_IV_Doppler_vs_OCTA_Macula}(d) are also indicated in Fig.~\ref{fig_IV_Doppler_vs_OCTA_Macula}(a), while those in Fig.~\ref{fig_IV_Doppler_vs_OCTA_Macula}(c) and \ref{fig_IV_Doppler_vs_OCTA_Macula}(d) are indicated in Fig.~\ref{fig_IV_Doppler_vs_OCTA_Macula}(b).

Although OCT images do not provide any quantitative flow measurements and no arteriovenous differentiation of vessels, they give accurate 3D information about the deep choroidal vasculature. Indeed the choroidal vessels leave very contrasted dark projections underneath them which helps locating them, and especially, the cross-sectional OCT images provide information that can help interpret the LDH panorama.

Fig.~\ref{fig_IV_Doppler_vs_OCTA_Macula}(a) allows to reconcile the information obtained with LDH, ICG-A and OCT. For example, in the bottom right of the image shown in Fig.~\ref{fig_III_Panorama_vs_ICG}(a), one can find the same choroidal veins than with ICG-A and LDH. On the bottom left, SS-OCT reveals the same choroidal veins as ICG-A, but not the choroidal arteries visible with LDH. In this Fig., we also show the remarkable positions of four choroidal arteries very noticeable with LDH in Fig.~\ref{fig_III_Panorama_vs_ICG}(a), and also visible with en-face SS-OCT in Fig.~\ref{fig_IV_Doppler_vs_OCTA_Macula}(a), and on the SD-OCT B-scan shown in Fig.~\ref{fig_IV_Doppler_vs_OCTA_Macula}(c). Although these arteries give a very strong signal with LDH, OCT data reveals that they are lying at the bottom of the choroid. This shows that the depth of field of the instrument allows to have both choroidal and retinal vessels simultaneously in focus despite the distance between them as measured with OCT being close to $400 \, \upmu \rm m$. Finally, the value of the information brought by LDH comparatively with OCT in this case is explicit, as whereas OCT does not provide clear information on the type of the detected choroidal vessels, as shown in subsection~\ref{subsection_HyperSpectral} in this same region of interest, LDH can undoubtedly identify these vessels as choroidal arteries.

In Fig.~\ref{fig_IV_Doppler_vs_OCTA_Macula}(b), we show an en-face SS-OCT image that was averaged over $70 \, \upmu \rm m$, approximately $30 \, \upmu \rm m$ below the choroid. At this depth, it is possible to see the entry points of some temporal distal SPCAs from the sclera into the choroid as they appear as dark spots (arrows) with SS-OCT. With LDH, these SPCAs can be detected as they carry a very large flow, but they appear as white and blurred spots, probably because of the strong intrasceral scattering. We marked with arrows some same SPCAs that were detected with LDH and SS-OCT in Fig.~\ref{fig_III_Panorama_vs_ICG}(a) and in Fig.~\ref{fig_IV_Doppler_vs_OCTA_Macula}(b), respectively. Additionally, we provide two examples of SD-OCT cross-section views of SPCAs identified with both LDH and SS-OCT in Fig.~\ref{fig_IV_Doppler_vs_OCTA_Macula}(e) and \ref{fig_IV_Doppler_vs_OCTA_Macula}(f). The course of these SPCAs is not straightforward as in both cases they are slightly tilted with respect to the optical axis. When looking closely at the LDH panorama in Fig.~\ref{fig_III_Panorama_vs_ICG}(a), it is possible to see how some of these SPCAs branch into choroidal arteries. For example, when following the course of the SPCA examined in Fig.~\ref{fig_IV_Doppler_vs_OCTA_Macula}(e), it is possible to see it branch into two large choroidal arteries.

Finally, in Fig.~\ref{fig_IV_Doppler_vs_OCTA_Macula}(d), we present a case of a choroidal artery making a U-turn to approach Bruch's membrane. We identified the same vessel with LDH where the upright part of the artery is very bright because the direction of its flow maximizes the Doppler sensitivity. Around this bright spot, it is possible to see with LDH this choroidal artery branch into smaller arteries in a vascular pattern that resembles a star. Although LDH is sensitive to lateral motion which allows to image vessels in en-face planes~\cite{Puyo2018}, the sensitivity of the instrument appears greater for flows with a direction parallel to the optical axis. This explain why the upright part of SPCAs and choroidal arteries approaching Bruch's membrane produce a greater power Doppler signal.

As choroidal vessels are entangled in multiple planes of the 3D choroid and the vertical summation of vessels makes it somewhat non trivial to interpret OCT images. The task of isolating any vessel in a particular en-face plane is complicated by the fact that averaging too many layers makes the resulting image unintelligible. However we have shown here that the information LDH and OCT can provide about the choroidal vasculature can complete one another to provide a better understanding of the overall choroidal anatomy. For example, we have demonstrated here that it is possible to follow a choroidal artery from the scleral SPCA to the moment where it branches into arterioles.

\subsection{Stages of the indocyanine green injection} \label{subsection_ICGA_phases}
\begin{figure}[t!]
\centering
\includegraphics[width = 1\linewidth]{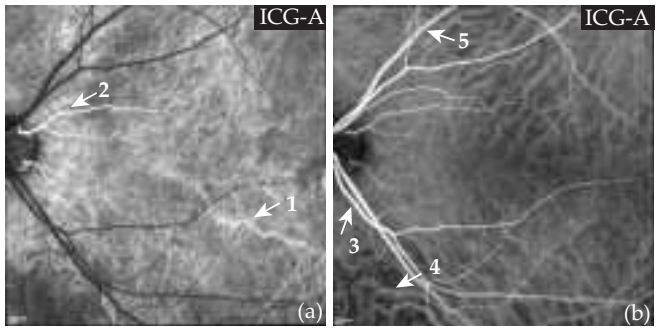}
\caption{Circulation of the ICG in the fundus vasculature. (a) Choroidal arteries and cilioretinal arteries are revealed by the contrast agent. (b) The contrast agent has reached the retinal arteries, the choroidal veins, and the retinal veins whereas choroidal arteries cannot be observed anymore. See \textcolor{blue}{\href{https://osapublishing.figshare.com/articles/media/ICG_video-angiography/7358855}{Visualization 1}} for the injection movie.
}
\label{fig_V_ICG_TwoPhases}
\end{figure}

ICG-A unfolds in several stages corresponding to the filling pattern of the dye inside the vascular circuit of the eye~\cite{Hayashi1985}, which gives a certain insight into the choroid and helps identify arteries and veins. We perfomed an ICG-A in the same subject on whom we made the power Doppler panorama presented in Fig.~\ref{fig_III_Panorama_vs_ICG} and we present in Fig.~\ref{fig_V_ICG_TwoPhases} the stages of the ICG circulation in the fundus vasculature that we observed experimentally. The vascular structures progressively appearing are marked by arrows. The general pattern is that following the natural course of blood through the vessels, the dye reveals arteries before veins, and the choroidal circuit is revealed slightly before the retinal circuit~\cite{Bottoni1994}. At the initial time of the injection (see \textcolor{blue}{\href{https://osapublishing.figshare.com/articles/media/ICG_video-angiography/7358855}{Visualization 1}}), no features are yet visible. Then Fig.~\ref{fig_V_ICG_TwoPhases}(a) shows that following the injection, the choroidal arteries and two cilioretinal arteries coursing towards the fovea are the first structures to be revealed. Subsequently the contrast agent reaches the retinal arteries, the choroidal veins, and finally the retinal veins as shown in Fig.~\ref{fig_V_ICG_TwoPhases}(b). At this stage, the cilioretinal arterial branches are still visible while the choroidal arteries no longer are.

Although they contribute to the retinal capillary bed like other retinal vessels, the cilioretinal arteries are branches from the PCAs. Consequently, they become visible concomitantly with choroidal arteries as they all originate from the PCA circulation. It can also be noticed that the first vessels revealed by ICG-A tend to be the ones that are the most visible with the LDH panorama shown in the previous section. For example, the large vessel marked by the arrow "1" in Fig.~\ref{fig_V_ICG_TwoPhases}(a) is also very visible in Fig.~\ref{fig_III_Panorama_vs_ICG}(a) which is another indication that high frequency power Doppler images preferentially reveal choroidal arteries.

With ICG video-angiography, a movie is recorded throughout the injection so the gain is set to a higher value to allow for a better temporal resolution, which diminishes the signal to noise ratio of the images. It is possible to obtain higher quality ICG angiograms of the choroidal arterial phase by using a lower gain. For example, analog photographic cameras have been very successfully used to image the choroidal arteries~\cite{Amalric1983}. However the choroidal arterial phase only lasts for a few seconds which makes the task challenging. This illustrates that ICG-A is a technique that is operator-dependent, especially when it is used with the intention to capture the moment when the arterial vasculature of the choroid is revealed.

\subsection{Power Doppler spectral analysis} \label{subsection_HyperSpectral}
\begin{figure}[t!]
\centering
\includegraphics[width = 1\linewidth]{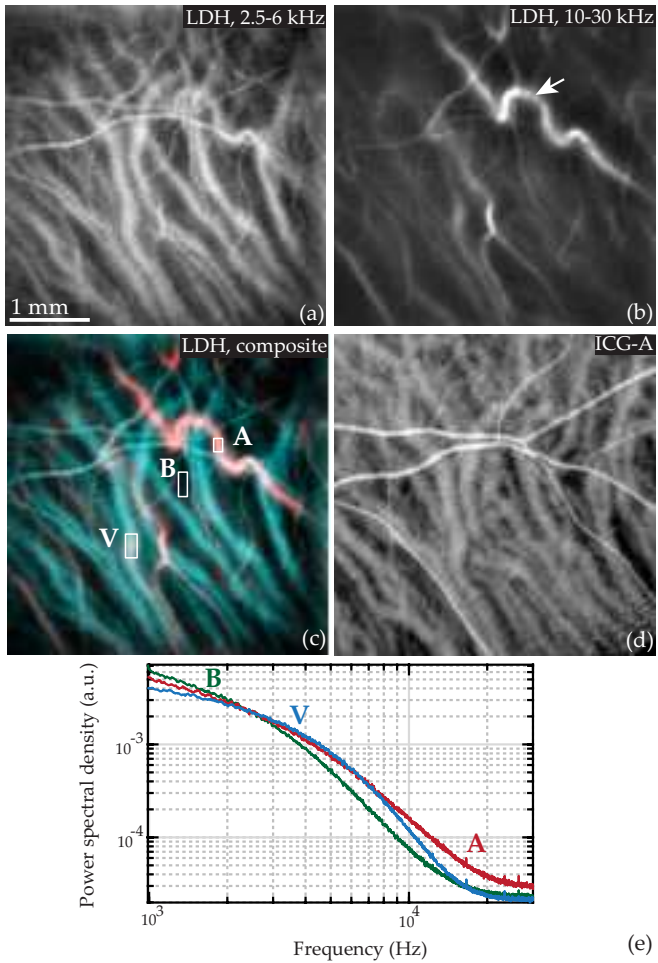}
\caption{Power Doppler spectral images and ICG-A. (a) Power Doppler image where the DPSD is integrated over $2.5-6 \, \rm  kHz$ which reveals vessels with smaller flows. (b) Power Doppler image integrated over $10-30 \, \rm  kHz$ revealing vessels with larger flows. (c) Composite color image of (a) and (b) encoded in the cyan and red channels, respectively. (d) ICG-A in the same region. (e) Power spectral density spatially averaged over the regions indicated in (c). See \textcolor{blue}{\href{https://osapublishing.figshare.com/articles/media/Laser_Doppler_holography/7358861}{Visualization 2}} for the power Doppler spectral movie.
}\label{fig_VI_HyperspectralImages_Spectra}
\end{figure}

\begin{figure}[t!]
\centering
\includegraphics[width = 1\linewidth]{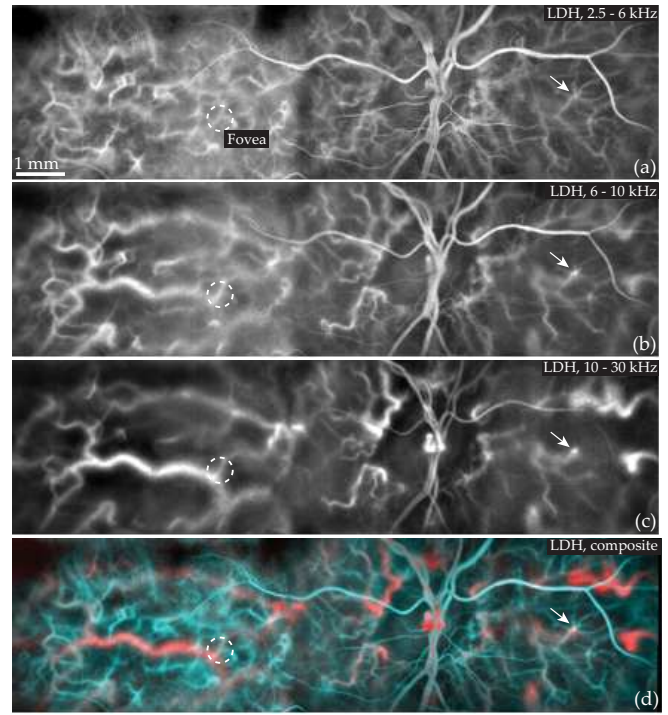}
\caption{Power Doppler spectral panoramas calculated using the frequency ranges (a)  $2.5-6 \, \rm  kHz$, (b) $6-10 \, \rm  kHz$, $10-30 \, \rm  kHz$. (d) Composite color panorama of (a) and (c) encoded in the cyan and red channels, respectively.
}\label{fig_9_HyperspectralImages}
\end{figure}

\begin{figure}[t!]
\centering
\includegraphics[width = 1\linewidth]{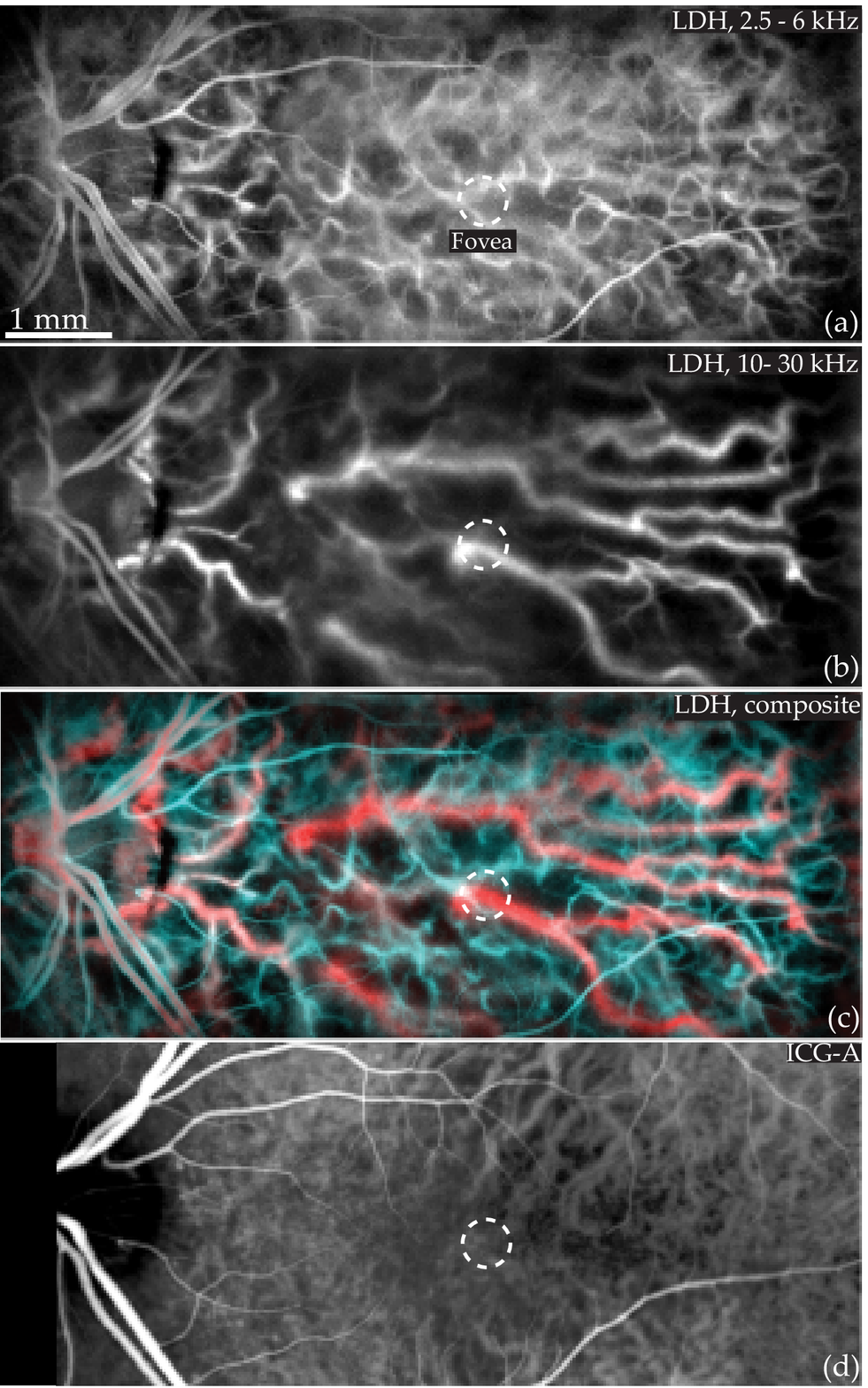}
\caption{Power Doppler spectral panoramas and ICG-A. (a) Low frequency ($2.5-6 \, \rm  kHz$) power Doppler panorama which reveals small flows. (b) High frequency ($10-30 \, \rm  kHz$) power Doppler panorama  revealing vessels with larger flows. (c) Composite color panorama of (a) and (b), encoded in the cyan and red channels, respectively. (d) ICG angiogram of the same region. See \textcolor{blue}{\href{https://osapublishing.figshare.com/articles/media/Laser_Doppler_holography/7358864}{Visualization 3}} for a power Doppler spectral movie in one of the area.
}\label{fig_VII_HyperspectralImages}
\end{figure}

We present here the results we obtained by using the process described in subsection~\ref{subsection_MethodsHyperspectrales}. We applied this process in a single power Doppler image in Fig.~\ref{fig_VI_HyperspectralImages_Spectra}, and in two stitched areas to demonstrate the reproducibility of the method. In Fig.~\ref{fig_9_HyperspectralImages}, we show power Doppler panoramas for the whole spectrum of the DPSD, and in Fig.~\ref{fig_VII_HyperspectralImages} we show the result of the process for part of the same dataset used in Fig.~\ref{fig_III_Panorama_vs_ICG} to show how the data can be processed otherwise to reveal different information about the choroid.

In Fig.~\ref{fig_VI_HyperspectralImages_Spectra}, the temporal Fourier transform was calculated over 8192 holograms (i.e. $0.14 \, \rm  s$), and two power Doppler images calculated for the frequency ranges $2.5-6 \, \rm  kHz$ and $10-30 \, \rm  kHz$ are shown in Fig.~\ref{fig_VI_HyperspectralImages_Spectra}(a) and \ref{fig_VI_HyperspectralImages_Spectra}(b), respectively. A composite color image was obtained by merging these two images in Fig.~\ref{fig_VI_HyperspectralImages_Spectra}(c) where low and high frequencies are encoded in the cyan and red channels, respectively. In Fig.~\ref{fig_VI_HyperspectralImages_Spectra}(d), an ICG angiogram of the same region of interest is shown. Finally in Fig.~\ref{fig_VI_HyperspectralImages_Spectra}(e) are plotted the DPSD measured in the regions of interest drawn in Fig.~\ref{fig_VI_HyperspectralImages_Spectra}(c). These spectra were calculated as the squared magnitude of the temporal Fourier transform of the reconstructed holograms amplitude.

Noticeably, the power Doppler images made with the frequency ranges $2.5-6 \, \rm  kHz$ and $10-30 \, \rm  kHz$ do not reveal exactly the same vessels. In this example, the $2.5-6 \, \rm  kHz$ power Doppler image is very similar to the ICG-angiogram, whereas the $10-30 \, \rm  kHz$ power Doppler image reveals vessels that cannot be observed with ICG-A; the largest vessel revealed with LDH (arrow) in Fig.~\ref{fig_VI_HyperspectralImages_Spectra}(b) even seems to leave a shadow on the ICG angiogram. The frequency range at which the vessels appear is not determined by the size of the vessels, as proven by the fact that the vessels revealed with the high frequency power Doppler image can be of both bigger or smaller size than those revealed by the low frequency power Doppler image. Figure~\ref{fig_VI_HyperspectralImages_Spectra}(c) offers an effective segregation of vessels based on the blood flow they carry: vessels appearing in red have a larger blood flow whereas vessels in blue have a lower blood flow. The spectra in Fig.~\ref{fig_VI_HyperspectralImages_Spectra}(e) were calculated in vessels of comparable sizes for "A" and "V" and in a third area "B" where although no vessels can be observed, there are unresolved retinal and choroidal capillaries. These spectra can be segmented into three parts. In the first range of frequency $0-2 \, \rm  kHz$, the DPSD in vessels "A" and "V" is smaller than that of the background "B". Then for $2-6 \, \rm  kHz$, the DPSD is greater in the area "V" and finally, for large frequencies such as $10-30 \, \rm  kHz$, the DPSD is clearly greater in the vessel "A".

This information is enough to determine that "A" and "V" actually mark an artery and a vein, respectively. First, whereas all the surrounding vessels were seen in the late phase ICG-angiogram, the "A" vessel was not observed, and as we saw in subsection~\ref{subsection_ICGA_phases}, this is typical of choroidal arteries. Secondly, this "A" vessel appears with very high Doppler frequency shifts which reveal very large flows, and choroidal arteries are known to have large flows. Other vessels appear at high Doppler frequency shifts despite being of a smaller caliber than the typical vessels appearing at low frequencies; these vessels were also not observed with ICG-A so we can assume they are choroidal arterioles that carry a comparable flow than choroidal veins of a greater size.

In Fig.~\ref{fig_9_HyperspectralImages}, power Doppler spectral images from five lateral adjacent locations were stitched following the process described in subsection~\ref{subsection_HuginStitchin} to reveal the choroidal vasculature over a wider field of view. Figure~\ref{fig_9_HyperspectralImages}(a-c) show power Doppler panoramas for the frequency ranges $2.5-6 \, \rm  kHz$, $6-10 \, \rm  kHz$, and $10-30 \, \rm  kHz$, respectively. Figure~\ref{fig_9_HyperspectralImages}(d) is the composite color panorama that was obtained by fusing power Doppler spectral images for the frequency ranges $2.5-6 \, \rm  kHz$ and $10-30 \, \rm  kHz$ (encoded in the cyan and red channels, respectively).

In Fig.~\ref{fig_9_HyperspectralImages}(a) is shown the power spectral image with the lowest frequency range that reveals the smallest detected flows. It is possible to observe large retinal vessels with the lumen of the vessels that appears dark because the local optical Doppler broadening occurs for higher frequency shifts than $2.5-6 \, \rm  kHz$. It is also possible to observe many choroidal vessels with small and intermediate sizes (see white arrow for example). In Fig.~\ref{fig_9_HyperspectralImages}(b), it is possible to notice choroidal arteries and retinal vessels of intermediate size are well visible. Finally, in Fig.~\ref{fig_9_HyperspectralImages}(c), it is possible to observe the largest choroidal arteries and other large choroidal structures temporally around the ONH with an improved contrast. The apparent diameter of retinal vessels is reduced when compared to the low frequency power Doppler image. Once again the color composite image in Fig.~\ref{fig_9_HyperspectralImages}(d) offers an effective segregation of vessels based on their blood flow that facilitates an arteriovenous differentiation of the choroidal vessels. The red vascular structures that stand out the most in Fig.~\ref{fig_9_HyperspectralImages}(d) are large choroidal arteries originating from temporal and nasal distal SPCAs and paraoptic SPCAs. Conversely, the vascular structures with a low blood flow that stand out in blue in Fig.~\ref{fig_9_HyperspectralImages}(d) are the small choroidal vessels.

In Fig.~\ref{fig_VII_HyperspectralImages}(a) and \ref{fig_VII_HyperspectralImages}(b), images from four lateral locations have been stitched for the frequency ranges $2.5-6 \, \rm  kHz$ and $10-30 \, \rm  kHz$, respectively. Figure~\ref{fig_VII_HyperspectralImages}(c) is the composite color panorama that was obtained by fusing power Doppler spectral images for low and high frequencies encoded in the cyan and red channels, respectively. Finally, Fig.~\ref{fig_VII_HyperspectralImages}(d) offers a comparison with ICG-A in the same region of interest.

In the low frequency power Doppler panorama shown in Fig.~\ref{fig_VII_HyperspectralImages}(a), large retinal vessels with their lumen in dark as well as smaller vessels can be observed. Especially, about 2 mm temporally from the fovea, a network of choroidal vessels of relatively small caliber (about $40 \, \upmu \rm m$) is visible. In Fig.~\ref{fig_VII_HyperspectralImages}(b), the small choroidal vessels cannot be observed anymore and instead very large choroidal vessels can be observed. Finally, Fig.~\ref{fig_VII_HyperspectralImages}(c) shows again a segregation of vessels based on their flow that leads to an arterioveinous differentiation: the large choroidal vessels that appear in red are arteries, whereas the large vessels in blue are mostly veins and the smaller ones are arterioles.

In the two power Doppler stitched images presented in Figs.~\ref{fig_9_HyperspectralImages} and ~\ref{fig_VII_HyperspectralImages}, we have shown that despite the high density of RPE and photoreceptor cells in the central retina, LDH allows us to reveal the underlying choroidal vasculature. Large choroidal arteries lying in the deepest layer of the choroid can be revealed and in both examples it could be noticed that the fovea lied right over a very large choroidal artery. Some other interesting vasculature structures revealed by LDH are the smaller choroidal vessels which have the size of choroidal arterioles or venules. These vessels are arterioles following this reasoning: as the choroidal veins are not visible at their starting points with LDH, choroidal venules should also not be observable with LDH as their flow is even lower. Interestingly, Fig.~\ref{fig_VII_HyperspectralImages}(d) shows that during the venous filling, ICG-A reveals neither the choroidal arterioles visible in the low frequency power Doppler image, nor the large choroidal arteries visible in the high frequency power Doppler image.

The comparative ICG angiogram we show in Fig.~\ref{fig_VII_HyperspectralImages} does not reveal any of these arterial choroidal structures of the macular region because the ICG angiogram is made at a late stage of the dye injection so choroidal arteries are no longer visible. However even in the early stage (cf Fig.~\ref{fig_V_ICG_TwoPhases}(b)), the structures revealed by LDH could here hardly be discerned with ICG video-angiography.

The power Doppler spectral analysis that we present here led to a clear discrimination of vessels according to their flows; in the posterior pole, this process allowed us to differentiate the choroidal arteries from the choroidal veins. The validity of this segregation is further discussed in section~\ref{sec_Discussion}.
\subsection{Vortex veins} \label{subsection_VortexVeins}

\begin{figure}[t!]
\centering
\includegraphics[width = 1\linewidth]{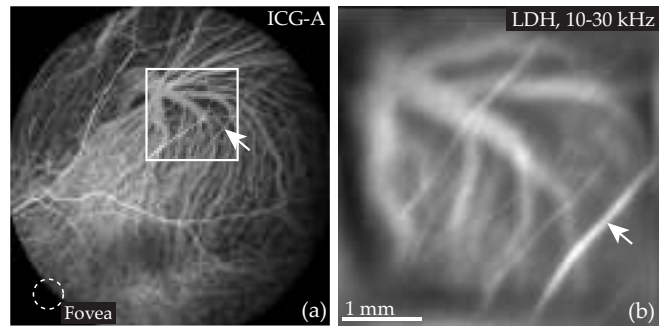}
\caption{Vortex vein imaged with ICG-A and LDH (a) ICG angiogram; the vortex vein is located at about 45 degrees of eccentricity; the fovea is indicated by the white circle and the rectangular box indicates the area covered by the power Doppler image (b) Power Doppler image integrated over the frequency range $10-30 \, \rm  kHz$. The arrow marks a choroidal artery visible with a bright contrast with LDH, and a darker contrast with ICG.
}\label{fig_VIII_VortexVein}
\end{figure}

The density of epithelial and photoreceptor cells decreases with the eccentricity from the fovea, so as these cells contribute to the scattering and absorption of light, a decrease in their density benefits the contrast of choroidal images. Imaging at a greater eccentricity seems to improve the visibility of choroidal vessels with both ICG-A and LDH. In the example presented in Fig.~\ref{fig_VIII_VortexVein}(b), LDH was used at about 45 degrees of eccentricity in the superior temporal choroid; the region of interest is delineated by the rectangle drawn in the ICG angiogram presented in Fig.~\ref{fig_VIII_VortexVein}(a). The power Doppler image, calculated by integrating the DPSD over $10-30 \, \rm  kHz$, reveals a vortex vein with a contrast and spatial resolution comparable to ICG-A. The arrow marks a choroidal artery visible with both ICG-A and LDH with a different contrast: with ICG-A this artery does not appear as bright as the surroundings vessels whereas with laser Doppler holography this artery appears brighter than the surroundings vessels. Considering these surroundings vessels are converging towards the vortex vein, they can be identified as choroidal veins.

In the previously shown example in Fig.~\ref{fig_VI_HyperspectralImages_Spectra}, the choroidal veins were visible for the frequency range $2.5-6 \, \rm  kHz$ whereas the veins close to the vortex veins are visible for the frequency range $10-30 \, \rm  kHz$. This large difference of flow can be explained by the fact that the flow in choroidal veins is expected to be considerably larger close to the vortex veins due to their large caliber and because they also drain the iris and ciliary body circulation which further increase the flow they carry. Conversely, in Fig.~\ref{fig_VII_HyperspectralImages}, some choroidal veins visible with ICG-A could not observed with LDH. It can be assumed that it is because in this region of interest the choroidal veins are too close to their starting points so their flow might be too weak and goes undetected by LDH.

This example shows that outside the posterior pole, ICG-A is able reveal choroidal arteries in the late injection phase, while on the other hand, choroidal veins become visible on high frequency power Doppler images. Nonetheless, the greater power Doppler signal of the artery compared to the other vessels despite its smaller diameter still allows us to identify it as a choroidal artery.

\section{Discussion and conclusion} \label{sec_Discussion}
In the present optical configuration, the depth of field of the instrument is large enough so that both retinal and choroidal vessels come into focus at the same reconstruction distance. This allows for the simultaneous visualization of flow in vessels from both of these vascular circuits. However, care should be taken when comparing flows between retinal and choroidal vessels as the layers lying in-between the retina and the choroid may have a non trivial effect on the spectrum of the transmitted light. Power Doppler images are calculated from the high-pass filtered DPSD: this quantity depends on both the number of Doppler shifted photons (amplitude of the spectrum) and their frequency shifts (spectral distribution of the energy). The photoreceptor and RPE layers between the retina and the choroid are known for their scattering and absorption properties which may alter the Doppler spectrum of the transmitted light. Consequently, although these layers possibly have only a negligible effect on the spectrum, without further investigation it is not possible to compare the power Doppler values for the retina and the choroid.

Still, the results we obtained are consistent with what can be expected: the flow rate through the retinal vessels is known to be significantly slower than the blood flow through the choroidal vasculature~\cite{AlmBill1973}. In our results, the signal of the retinal arteries lies in a lower frequency range than the frequency range for the choroidal arteries (cf Figs.~\ref{fig_VI_HyperspectralImages_Spectra}, \ref{fig_9_HyperspectralImages}, \ref{fig_VII_HyperspectralImages}). Additionally, comparisons of flow inter-choroidal vessels should be legitimate as the light backscattered by these structures goes through the same layers. Finally and most importantly, although the results are not quantitative for velocities, they still allow for a visualization of the architecture of the choroidal vasculature.

Recording interferograms with a high-speed camera is essential to the purpose of LDH. Wideband measurements must be performed with a Nyquist-Shannon frequency above the frequency of the signal to be sampled~\cite{Puyo2018}. As shown in the spectra calculated in Fig.~\ref{fig_VI_HyperspectralImages_Spectra}, the Doppler broadening in choroidal arteries can reach $30 \, \rm kHz$ or more (as visible in \textcolor{blue}{\href{https://osapublishing.figshare.com/articles/media/Laser_Doppler_holography/7358861}{Visualization 2}} and \textcolor{blue}{\href{https://osapublishing.figshare.com/articles/media/Laser_Doppler_holography/7358864}{Visualization 3}}), and a major part of the Doppler signal for these vessels with a large flow lies in frequencies above $10 \, \rm kHz$. Consequently, in this configuration it is necessary to be working with a camera that has a framerate of several tens of kHz to obtain a better signal-to-noise ratio. Using a powerful reference beam allows to dramatically increase the framerate in comparison with optical configuration where the images on the camera are formed only with the light backscattered by the retina, which is why digital holography is convenient to make wideband laser Doppler measurements.
We had previously shown that the pulsatile signal in retinal blood vessels can be separated from the signal of global eye motion~\cite{Puyo2018}. In the case of choroidal vessels, not only the Fourier filtering helps removing the Doppler contribution of global eye motion, but it also helps to select the photons backscattered to the sensor despite the strongly scattering layers lying in-between. Finally, as was also discussed in~\cite{Puyo2018}, working with highly Doppler shifted photons also means working with multiply scattered light, which allows the angiographic contrast to be sensitive to lateral motion. This property of power Doppler images is yet again ascertainable as in all the images presented in this manuscript, the vast majority of the vessels are oriented perpendicularly to the optical axis.

As demonstrated in subsection~\ref{subsection_HyperSpectral}, a power Doppler spectral analysis allows to discriminate vessels according to the intensity of their flow. In Fig.~\ref{fig_VI_HyperspectralImages_Spectra} this approach allowed to reveal choroidal veins with a low frequency power Doppler image while large choroidal arteries were revealed with a high frequency power Doppler image. In the other example considered in Fig.~\ref{fig_VII_HyperspectralImages}, choroidal arteries can be seen once again at high frequencies, but on the other hand choroidal arterioles can be seen at low frequencies with choroidal veins. And finally, in the example presented in Fig.~\ref{fig_VIII_VortexVein}, choroidal veins near a large vortex vein can be seen at high Doppler frequency shifts. Thus, the separation that can be made with this spectral analysis is based on the flow of the vessels and not whether they are arteries or veins. However we identified with certainty as arteries some vessels such as those that appear at high frequencies in the macular region for several reasons: (i) they appear at an early stage of ICG-A, (ii) they can be seen with LDH for strong Doppler frequency shifts, and (iii) they have a geometry consistent with the expected disposition of choroidal arteries.

Because the fovea has the highest needs for oxygen and nutrients and because numerous ocular diseases tend to form primarily in this area, the choroidal vasculature in the submacular region is of paramount importance. We have shown that comparatively with other ophthalmic instruments, LDH gives new insights about choroid, especially in the submacular area. Even operating at $785 \, \rm nm$ LDH can provide high contrast imaging of choroidal vessels in human, with a resolution at least equivalent to state of the art ICG-A and OCT instruments. LDH preferentially reveals the choroidal arterial structures due to their greater blood flow, and provides the opportunity to discriminate arteries from veins in the posterior pole of the choroid based on a power Doppler spectral analysis. Combined with the accurate volumetric information accessible with OCT, LDH offers the possibility to gain major insights about the physiology of the choroid non-invasively.

\section*{Funding}
This work was supported by LABEX WIFI (Laboratory of Excellence ANR-10-LABX-24) within the French Program Investments for the Future under Reference ANR-10-IDEX-0001-02 PSL, and the European Research Council (ERC Synergy HELMHOLTZ, grant agreement \#610110). The Titan Xp used for this research was donated by the NVIDIA Corporation.

\section*{Acknowledgments}
The authors would like to thank C\'eline Chaumette, Carlo Lavia, and Alain Gaudric for data acquisition and helpful discussions, and Kate Grieve for advice and language corrections.

\section*{Disclosures}
The authors declare that there are no conflicts of interest related to this article.

\section*{Supplementary Material}
\noindent
\textcolor{blue}{\href{https://youtu.be/s-_LuNT2cuQ}{Supplementary Visualization 1}}. \newline

\bibliography{./Bibliography}

\end{document}